\documentclass[11pt,a4paper]{article}
\usepackage{jinstpub}

\title{
Silicon $\mathbf{4\pi}$ spectrometer for $\mathbf{\beta}$\nobreakdash-decay electrons with energies up to 3~MeV.\\
\vspace{5pt}
\small
}

\author[b]{I.E.~Alexeev}
\author[a]{S.V.~Bakhlanov}
\author[a]{E.A.~Chmel}
\author[a]{A.V.~Derbin}
\author[a]{I.S.~Drachnev}
\author[a]{I.M.~Kotina}
\author[a]{M.S.~Mikulich}
\author[a]{V.N.~Muratova}
\author[a]{N.V.~Niyazova}
\author[a]{D.A.~Semenov}
\author[a]{M.V.~Trushin}
\author[a]{E.V.~Unzhakov}

\affiliation[a]{Petersburg Nuclear Physics Institute, NRC Kurchatov Institute, Gatchina, Russia}
\affiliation[b]{V.~G.~Khlopin Radium Institute, St.~Petersburg, Russia}

\abstract{
    We present a description of the originally developed $\beta$\nobreakdash-spectrometer consisting of two Si{(Li)}\nobreakdash-detectors with sensitive area thickness above 8~mm and $4\pi$\nobreakdash-geometry.
    The full absorption spectrometer allows for direct measurements of $\beta$\nobreakdash-spectra, disregarding the corrections to the response function induced by the electron backscattering from the crystal surface.
    In case of $\beta$\nobreakdash-spectra of transitions to the excited state of the daughter isotope additional $3^{\prime\prime}$ BGO\nobreakdash-detector is used in order to detect the $\gamma$\nobreakdash-quanta in coincidence with the pair of Si{(Li)}\nobreakdash-spectrometers.
}


\begin{document}
\maketitle

\section{Introduction}\label{sec:intro}
A high precision knowledge of $\beta$-spectrum shape of particular isotopes is absolutely necessary for determination of the neutrino oscillation parameters in experiments with reactors and artificial neutrino sources.
The positive outcome of the LSND experiment~\cite{Athanassopoulos1998} obtained almost two decades ago, the calibration results of the radio-chemical Ga-Ge solar neutrino detectors~\cite{Abdurashitov1999,Hampel1999} and the recent calculations of the reactor neutrino spectrum known as the ``reactor anomaly''~\cite{Mention2011,Giunti2022} have raised a question about existence of the sterile neutrino, which is mainly connected with the neutrino mass state shifted from the three known states by $\delta m_{14}^2 \sim 1$~eV and with effective mixing angle $\sin^2(2\theta_s) \sim 0.1$.
In addition to the series of reactor and accelerator experiments looking for neutrino oscillations into the sterile state, there are projects employing artificial neutrino sources.
The most popular electron neutrino source is $^{51}$Ce isotope, proposed for use with various types of neutrino detectors and actually used for mentioned calibrations of SAGE and Gallex/GNO and BEST experiment~\cite{Barinov2022a}.

Among the artificial sources of electron antineutrinos, $^{144}$Ce\nobreakdash-$^{144}$Pr is the most promising one, which was planned for use with KamLand~\cite{Gando2013} and Borexino~\cite{Bellini2013} detectors.
Unfortunately, the most advanced project of Borexino SOX was canceled for several reasons, including the ones unrelated to the technical side of the project.
One of the particular objectives to be solved in any experiment with $^{144}$Ce\nobreakdash-$^{144}$Pr is the precision measurement of its $\beta$\nobreakdash-spectrum, aimed to determine the intensity and shape of the antineutrino spectrum.
This task is of the utmost importance, since calculated sensitivity of the experiment to the oscillation parameters $\delta m_{14}^2$ and $\sin^2(2\theta_s)$ can be achieved only as long as the ratio between heat power and activity of the source and the expected rate of inverse $\beta$\nobreakdash-decay are known with required precision ($\leq 1.5$\% in case of Borexino SOX).

The measurements of $^{144}$Ce\nobreakdash-$^{144}$Pr and $^{210}\rm{Bi}$ $\beta$\nobreakdash-spectra via the silicon semiconductor $\beta$\nobreakdash-spectrometer in classic ``target-detector'' layout has been carried out previously in~\cite{Bazlov2018,Alexeev2018,Alekseev2020}.
The isotope $^{210}\rm{Bi}$ belongs to the natural radioactive decay chain of $^{238}\rm{U}$ and as a product of the radioactive gas $^{222}\rm{Rn}$ presents inside or on the surface of almost all structural materials.
The precise measurement of $^{210}\rm{Bi}$ $\beta$\nobreakdash-spectrum  remains a crucial task for background modeling of modern neutrino detectors, as well as for the dark matter
searches or other low-background experiments.

Current paper contains the description of the custom made $4\pi$ $\beta$\nobreakdash-spectrometer with $4\pi$\nobreakdash-geometry based on Si(Li) semiconductor detectors.
The designed spectrometer is capable of registering electron's full energy, thus solving the problem of crystal surface backscattering, whose probability can reach tens of percent, depending on the electron energy and incidence angle.
The initial results of $^{144}$Ce\nobreakdash-$^{144}$Pr and $^{210}\rm{Bi}$ $\beta$\nobreakdash-spectra measurements via the $4\pi~\beta$\-spectrometer consisting pair of Si(Li) detectors were published in~\cite{Bakhlanov2019,Alekseev2021,Alekseev2021a,Alekseev2021b}.

\section{Experimental setup}\label{sec:setup}
Silicon Lithium (Si(Li) - lithium doped silicon) detectors were fabricated from a single-crystal p\nobreakdash-type silicon with specific resistivity of $4$~k$\Omega$/cm and mean carrier lifetime of $800$~$\mu$s, using a standard well-proven process, developed at Petersburg Nuclear Physics Institute NRC KI\@.
Detailed specifications for such detectors may be found in~\cite{Bazlov2018,Alexeev2018,Bazlov2019}.

\begin{figure}[b]
    \centering
    \includegraphics[width=.8\linewidth]{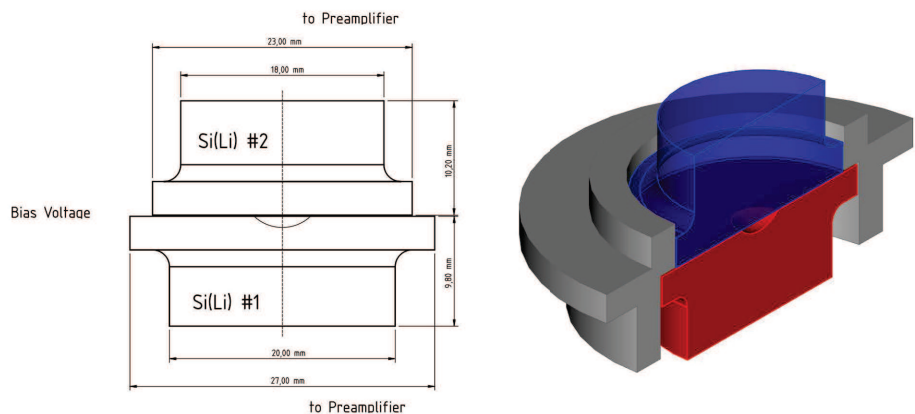}
    \caption{\textbf{Left:} The layout of $\beta$-spectrometer with two planar Si(Li) detectors. \textbf{Right:} Vertical cross-section of spectrometer assembly with aluminum casing.}\label{fig:setup_layout}
\end{figure}

Two Si(Li) detectors with sensitive area thickness above $8$~mm were custom-designed specifically for this experiment.
Detector bulks were made in a ``mushroom'' shape with outer diameters of $27$~mm  and $23$~mm, sensitive area diameters of $20$~mm and $18$~mm, correspondingly. Both detectors had height of $9$~mm.
Different outer diameters were dictated by the dimensions of casing necessary for convenient face to face attachment of detectors (see Fig.~\ref{fig:setup_layout}).

\begin{figure}
    \centering
    \includegraphics[width=.8\linewidth]{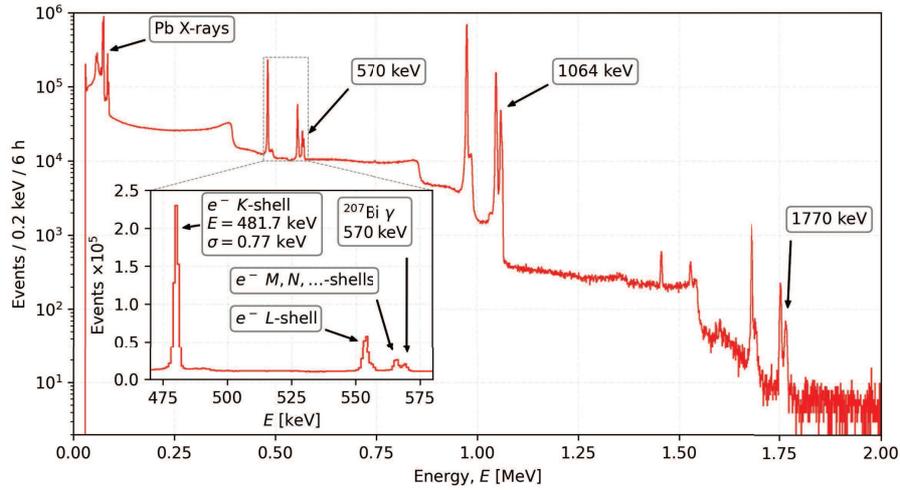}
    \caption{The calibration spectra of the Si(Li) detector measured with internal $^{207}$Bi. The inset shows the electron peaks corresponding to internal conversion from K, L, M, and N shells of the 570 keV nuclear level.}
    \label{fig:sp_207Bi_int}
\end{figure}

Performance of the manufactured detectors was tested in a separate vacuum cryostat, using gamma-rays (X-rays and nuclear lines) and electrons (conversion and Auger) produced by the $^{207}$Bi source (see Fig.~\ref{fig:sp_207Bi_int}).
The energy resolution of both detectors was found to be $\rm{FWHM} \approx 2.0$~keV for $482$~keV electrons.
The determined thickness of the dead layer, contributed by Pd and Au coatings and surface Si layer turned out to be $\sim 500$~nm in silicon equivalent.
Upon traversing this depth, an electron with energy of $20$~keV or $3$~MeV would lose around $1$~keV or $0.1$~keV, correspondingly.

The comparison of measured X-ray intensities of K$_{\alpha 1}$, K$_{\alpha 2}$, K$_{\beta 123}$ Pb peaks against Geant4 simulations allowed us to evaluate the thickness of detector's sensitive region.
Thus determined thicknesses for both detectors were found to be above $8.5$~mm, providing the full absorption for electrons with energies up to $3.3$~MeV, disregarding the backscattering cases.
A small cavity was machined in the middle part of a single detector, with diameter of $5$~mm and $1$~mm deep.
The investigated $\beta$\nobreakdash-source ($^{144}$Ce, $^{210}$Pb) was deposited inside this cavity, directly upon the golden coating of the planar Si(Li) detector.
The second detector was placed, without any gap, on top of the cavity and the bias voltage was supplied to the shared $n^+$-contact~(Fig.~\ref{fig:setup_layout}).

The whole composition was placed inside the vacuum cryostat and cooled down to the liquid nitrogen temperature.
The refrigeration is usually necessary for reduction of the detector reverse current and consequent improvement of the energy resolution.
Both Si(Li) detectors had their own charge sensitive preamplifier with resistive feedback and cooled first field effect transistor (FET)\@.
Two separate sets of electronics were used for the signal processing: CAMAC and VME\@.
CAMAC channel consisted of the BUI-3K amplifier and 12-bit successive approximation ADC with 4096 channels.
The coincidence circuit was implemented at the ADC level via the conversion-blocking signals.
The signals in VME channel were digitized by 14-bit digitizer module (CAEN V1725) with 8 channels and $250$~MHz sampling rate.
Operation via two separate electronic sets allows for a comparison of coincidence selection efficiency and energy resolution between analogue and digital channels.

Coincidence circuit includes $3^{''}$ BGO scintillation detector for selection of events of $\beta$\nobreakdash-decay into the excited nuclear states.
The choice of relatively small BGO-detector was driven by requirement for high $\gamma$-registration efficiency in case of low background of random coincidences.
BGO-detector was placed at $25$~mm distance form the shared surface of Si(Li) detectors (i.~e.\ from the $^{144}\rm{Ce}-^{144}\rm{Pr}$ source), providing $20$\% geometric efficiency.
The data produced by registered events is represented as a series of time and amplitude signals from two Si(Li) and single BGO detectors.
The entire assembly of Si(Li) and BGO detectors was surrounded by moderate layer of passive shielding ($\sim 30$~g/cm$^2$) in order to reduce the natural radioactivity background.

The adopted (anti-)coincidence operation mode allows for the direct measurement of $\beta$\nobreakdash-spectra, which does not require any complex corrections describing the electron backscattering from the surface of the detector.
The deviation of response function from pure Gaussian is related mostly to the energy loss of the electron inside the source material and dead layer of the detector and also with escape of electron bremsstrahlung from the detector volume.

The fabrication of ultra thin homogeneous $\beta$\nobreakdash-sources is a whole separate topic \cite{Alekseev2020}, far too complex to fit inside this article.
We would like to note, though, that presence of low additional $\alpha$\nobreakdash-activity in the source helps to ensure the thickness and density homogeneity of the source.

The contribution of total radiative losses increases along with the electron energy.
In case of silicon, for $1$~MeV and $3$~MeV electrons they come out at $0.8$\% and $4.7$\%, correspondingly.
Differential radiative loss $dE/dx$ also rises accordingly from $1.5$\% to $4.7$\%.
The correction that takes bremsstrahlung escape probability into account can be obtained through the numerical simulations with use of BGO coincidence spectrum.

\section{Results}\label{sec:results}
\begin{figure}
    \centering
    \includegraphics[width=.8\linewidth]{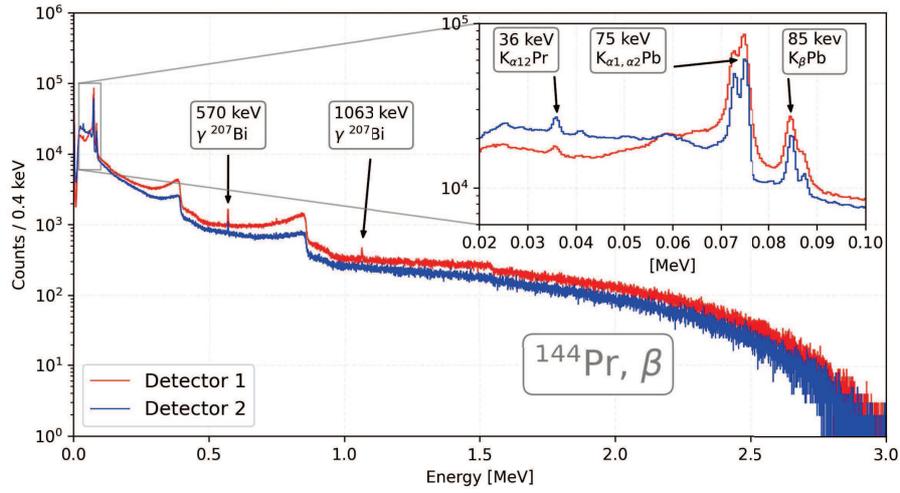}
    \caption{The calibration spectra of the Si(Li) detectors measured with external $^{207}$Bi (spectrum of detector with cavity is represented by red line). The inset shows zoom-in of the low-energy region.}\label{fig:spec_sili_calib}
\end{figure}

After installation of the Si(Li) detector assembly into the cryostat, the energy calibration was performed via external $^{207}$Bi source that was placed onto the cryostat beryllium window.
While the calibration was performed with $^{144}\rm{Ce}-^{144}\rm{Pr}$ source present, the spectrum clearly showed X-ray ($75 - 78$~keV) and $\gamma$\nobreakdash-line ($570$~keV, $1063$~keV) full absorption peaks, and also the prominent Compton scattering cutoff (see Fig.~\ref{fig:spec_sili_calib}).
These spectral features were used to calibrate both detectors.

\begin{figure}
    \centering
    \includegraphics[width=.8\linewidth]{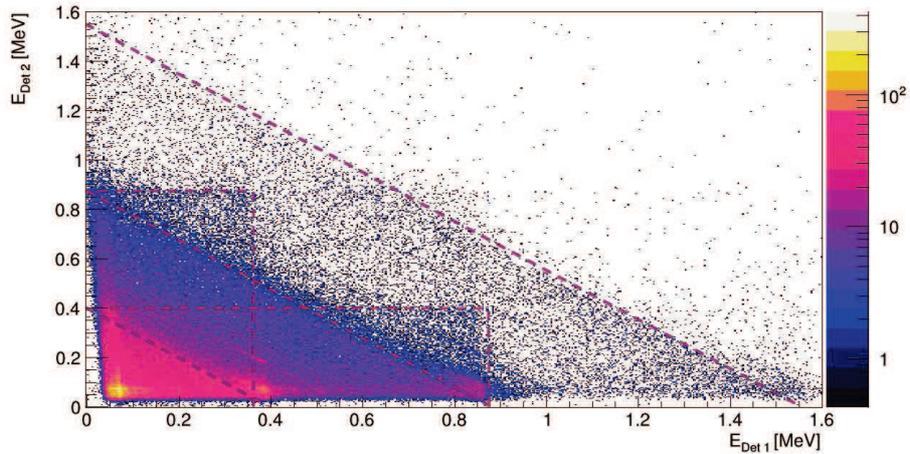}
    \caption{The coincidence spectrum of two combined Si(Li) detectors measured with $^{207}$Bi source. Slanted lines correspond to the Compton scattering of $570$~keV, $1064$~keV and $1770$~keV $\gamma$-quanta. Rectangles denote events from the $\gamma$-cascade $1064$~keV  and $570$~keV.}
    \label{fig:spec_sili_coinc}
\end{figure}

Coincidence spectrum of Si(Li) detectors with external $^{207}$Bi source  is given in Fig.~\ref{fig:spec_sili_coinc}.
Slanted lines represent the total Compton cutoff energies of $570$~keV, $1063$~keV and $1779$~keV $\gamma$-lines.
Maximum electron energy in case of reverse scattering amounts to $E_C = 2 E_\gamma^2 / (2 E_\gamma + m_e)$, yielding $394$~keV, $857$~keV and $1547$~keV, correspondingly.
The coincidences of signals from Si(Li) detectors are mostly caused by Compton electrons, traversing from upper detector into the bottom one.
Rectangles in the Fig.~\ref{fig:spec_sili_coinc} denote the events when cascaded $1064$~keV  and $570$~keV $\gamma$-quanta undergo Compton scattering in different detectors.

\begin{figure}
    \centering
    \includegraphics[width=.8\linewidth]{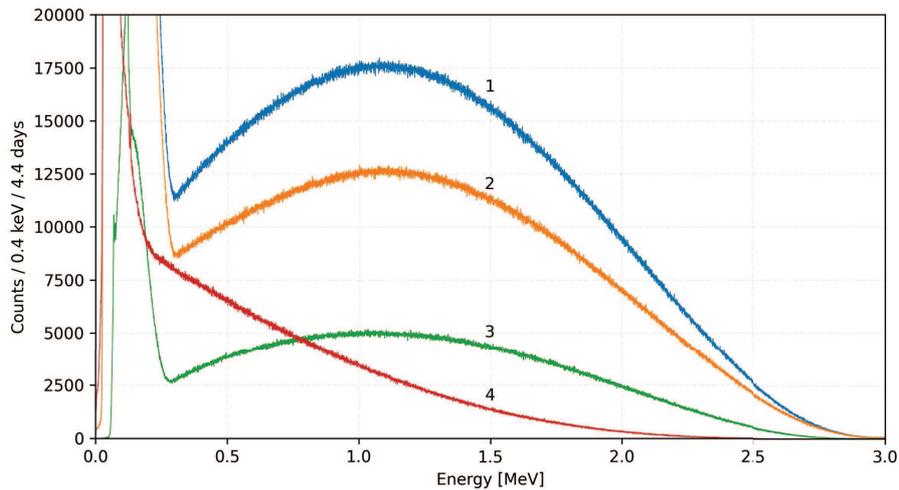}
    \caption{The spectra of $^{144}$Ce-$^{144}$Pr source measured by $4\pi$ $\beta$-spectrometer: 1 -- spectrum of total energy of all registered events, 2 -- energy spectrum of single-site events, 3 -- spectrum of total energy of multi-site events, 4 -- single detector energy spectrum of multi-site events.}\label{fig:spec_total}
\end{figure}

The studied electron spectra produced by $^{144}$Ce-$^{144}$Pr $\beta$\nobreakdash-decays registered by a single or both Si(Li) detectors are shown in the Fig. \ref{fig:spec_total}.
Comparing curves 1 and 3 one can find that the share of electrons backscattered by detector surface at $1$~MeV energy amounted to $28$\%.
This value decreases down to $20$\% as electron energy rises up to $2.5$~MeV and is determined by the particular geometry of the source.
It is worth noting that in case of backscattering the shape of registered energy spectrum (Fig.~\ref{fig:spec_total}, curve~4) significantly varies from the actual electron energy spectrum (Fig.~\ref{fig:spec_total}, curve~1).
Full $\beta$-spectrum 1 obtained as a sum of spectra 2 and 3, solves the problem of determining the response function with respect to electron backscattering.

Provided the electron spectrum was measured with almost Gaussian response function, the antineutrino spectrum can be obtained from the relation $Q_\beta = E_e + E_\nu$, here $Q_\beta$ is the $\beta$-spectrum end-point energy, $E_e$ and $E_\nu$ - energies of electron and antineutrino, correspondingly.
In case of $^{144}\rm{Ce}-^{144}\rm{Pr}$ source it is possible for neutrino energies below $2.7$~MeV (difference between $\beta$\nobreakdash-spectrum end-point energies  of $^{144}$Pr and $^{144}$Ce).
At energies below $300$~keV the measured electron spectrum becomes a superposition of $^{144}$Pr and $^{144}$Ce $\beta$-spectra and the antineutrino spectrum must be determined using the theoretical corrections and shape factor parameters, obtained from the fitting of $(0.3 - 3.0)$~MeV interval of $^{144}$Pr $\beta$-spectrum.

The measurement of Si(Li) detector spectra in coincidence with signals from BGO detector allows for separation of $\beta$\nobreakdash-spectra corresponding to transitions that produce daughter nuclei in the excited states.
This appears to be important aspect of the measurements in case of $^{144}$Pr isotope, since allowed $0^- \rightarrow 1^-$ transition from ground state $0^-$ of $^{144}$Pr into $1^-$ $2186$~keV level of $^{144}$Nd has well defined shape and does not require introduction of shape factor function for fitting.
The agreement of the total measured spectrum shape to the shape of the allowed $\beta$\nobreakdash-transition represents an important validity criterion for the measurements, response function and fitting procedure.

\begin{figure}[t]
    \centering
    \includegraphics[width=.8\linewidth]{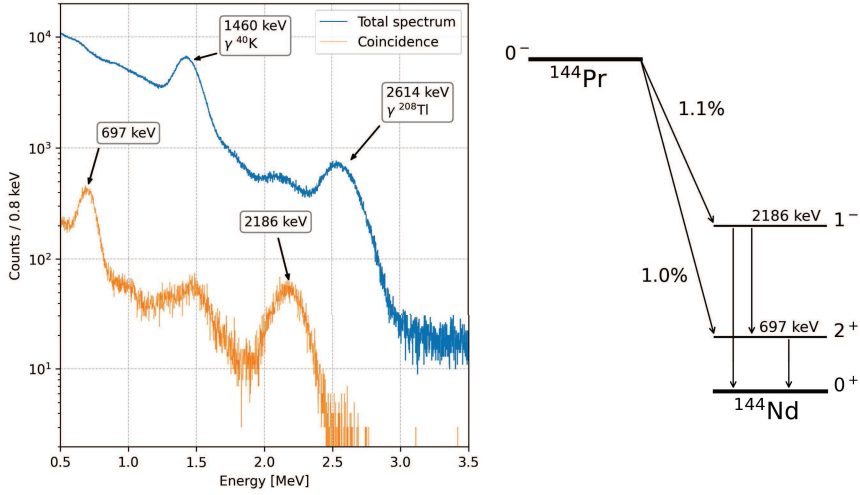}
    \caption{\textbf{Left:} Spectrum of events acquired with BGO detector (total and in coincidence with Si(Li) spectrometer). \textbf{Right:} Decay scheme for $^{144}$Pr $\beta$-decay.}\label{fig:spec_bgo}
\end{figure}

The $1^-$ $2186$~keV nuclear level of $^{144}$Nd can discharge directly into the ground state or through the intermediate $2^+$ state, emitting $2186$~keV ($0.69$\%) or $1489$~keV and $697$~keV ($0.28$\%) $\gamma$\nobreakdash-quanta, as shown on the inset of Fig.~\ref{fig:spec_bgo}.
The same figure presents the total energy spectrum of events registered by BGO detector and the spectrum of coincidence with $\beta$\nobreakdash-spectrometer events.
The full spectrum contains prominent full absorption peaks of $1460$~keV ($^{40}$K) and $2614$~keV ($^{208}$Tl) $\gamma$\nobreakdash-lines from the natural chain of $^{232}$Th.

The coincidence spectrum (Fig.~\ref{fig:spec_bgo}, curve~2) contains $697$~keV and $2186$~keV peaks, which correspond to the $\beta$\nobreakdash-transitions into the excited states of $^{144}$Nd nucleus.
In order to select transitions to the $1^-$ state, we set the condition on the energy of $\gamma$\nobreakdash-quantum registered by BGO detector to be above $1$~MeV.
This condition is fulfilled for the $\gamma$-quanta from the full absorption peak and part of the Compton scattering tail and allows for exclusion of transitions to $2^+$ state followed by a single $697$~keV $\gamma$-quantum.

\begin{figure}[t]
    \centering
    \includegraphics[width=.8\linewidth]{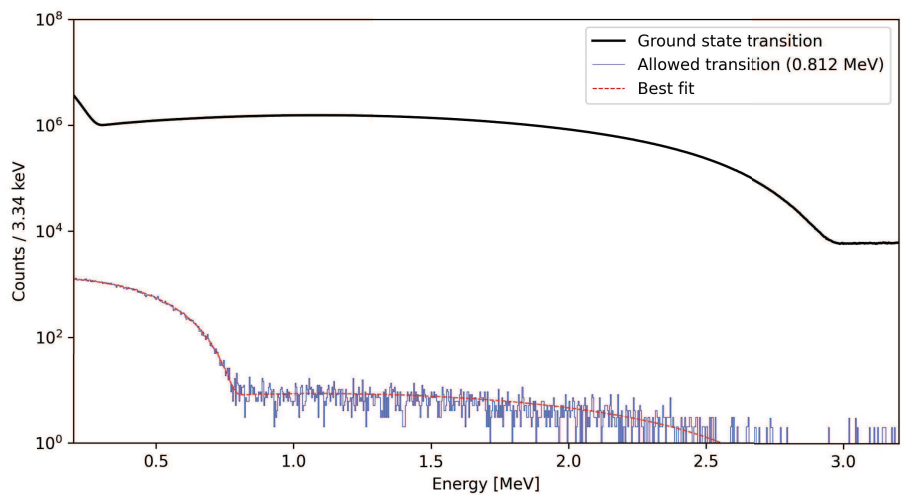}
    \caption{The spectrum of allowed $\beta$-transition ${^{144}}\mathrm{Pr} (0^-) \rightarrow {^{144}\mathrm{Nd} (1^-)}$ ($Q_\beta = 0.812$~MeV), measured in coincidence with BGO detector signal in comparison against the transition to ground state $^{144}\rm{Nd}~(0^+)$ ($Q_\beta = 3.0$~MeV). Fit result is shown by dashed red line.}\label{fig:spec_allowed}
\end{figure}

The spectrum of Si(Li) detectors in coincidence with BGO signals that exceeded $1$~MeV is given in Fig.~\ref{fig:spec_allowed}.
It consists of the $\beta$\nobreakdash-spectrum of the allowed transition of $^{144}\rm{Pr}~(0^-)$ to the excited state of $^{144}\rm{Nd}~(1^-)$ and random coincidences.
The spectrum shape is in good accordance with theoretical shape of the allowed $\beta$\nobreakdash-transition with $812$~keV endpoint, thus confirming that the response function of the $4\pi$ $\beta$\nobreakdash-spectrometer is indeed very similar to Gaussian.
The same Fig.~\ref{fig:spec_allowed} shows the full spectrum, mainly attributed to the ground state transition ${^{144}}\rm{Pr}~(0^-) \rightarrow ^{144}\rm{Nd}~(0^+)$.

\section{Conclusion}\label{sec:conclusion}
We presented the performance review of the newly developed spectrometer suited for the measurements of $\beta$\nobreakdash-spectra with endpoint energies below $3$~MeV.
The spectrometer consists of two Si(Li) detectors with thickness of the sensitive layer above $8$~mm and a BGO scintillation crystal for studying of transitions to the excited states.
The detector response function function proved to be very similar to Gaussian and does not contain the term caused by the electron backscattering form the crystal surface.
This feature allows for the direct measurement of $\beta$\nobreakdash-electron energy and, consequently, obtaining the electron antineutrino spectrum.
In combination with the scintillation BGO detector, the spectrometer can be employed for acquisition of spectra that correspond to the transitions to the excited states of the daughter nuclei.
The designed spectrometer was successfully employed for measurements of $\beta$\nobreakdash-spectra of $^{144}$Ce-$^{144}$Pr and $^{210}$Bi sources \cite{Bakhlanov2019,Alekseev2021,Alekseev2021a,Alekseev2021b}.

This work has been carried out with support of Russian Foundation for Basic Research (projects \#19\nobreakdash-02\nobreakdash-00097, \#20\nobreakdash-02\nobreakdash-00571) and Russian Scientific Foundation (projects \#21\nobreakdash-12\nobreakdash-00063, \#22\nobreakdash-22\nobreakdash-00017).

\bibliographystyle{unsrt}
\bibliography{ref}

\end{document}